# Ultrafast spin dynamics in the proximate quantum spin liquid $\alpha$-$RuCl_3$

Haochen Zhang[1,2], Subin Kim[2], Young-June Kim[2], Hae-Young Kee[2,3], and Luyi Yang[1,2,4,5,*]

[1]State Key Laboratory of Low Dimensional Quantum Physics, Department of Physics, Tsinghua University, Beijing 100084, China

[2]Department of Physics, University of Toronto, Toronto, Ontario M5S 1A7, Canada

[3]Canadian Institute for Advanced Research, Quantum Materials Program, Toronto, Ontario M5G 1M1, Canada

[4]Frontier Science Center for Quantum Information, Beijing 100084, China

[5]Collaborative Innovation Center of Quantum Matter, Beijing 100084, China

[*]e-mail: luyi-yang@mail.tsinghua.edu.cn

## Abstract

$\alpha$-$RuCl_3$ is a Kitaev material suggested to be a proximate quantum spin liquid in a certain temperature and magnetic field range. Nonequilibrium measurements of transient dynamics have been proposed to detect fractionalized particles that emerge in the spin liquid and to possibly drive the system into novel photoinduced magnetic states that cannot be accessed by conventional equilibrium probes. Here we study ultrafast spin dynamics of photoinduced excitations in $\alpha$-$RuCl_3$ using pump-probe transient grating spectroscopy. In the real part of the complex transient reflectance change $\Delta R/R$, we observe the long-range antiferromagnetic correlation near the Néel temperature. Most intriguingly, above the Néel temperature in the Kitaev paramagnetic phase, we reveal a photoexcitation component sensitive to the in-plane magnetic field in the imaginary part of $\Delta R/R$. This component exhibits two distinct lifetimes of about tens of picoseconds. This photoexcitation component may be connected to novel photoexcited states in the Kitaev quantum spin liquid, and its lifetimes likely reflect the dynamics of unconventional spin excitations in the Kitaev model.

## Introduction

Quantum spin liquids are an exotic state of matter with strongly entangled spins that lack magnetic order[1–3]. Kitaev materials[4,5], such as $Na_2IrO_3$ (Ref.[6]), polytypes of $Li_2IrO_3$ (Refs.[7–10]), $H_3LiIr_2O_6$ (Ref.[11]), and especially $\alpha$-$RuCl_3$ (Ref.[12]) (hereafter $RuCl_3$), are the most promising candidates for realizing a quantum spin liquid state.

In $RuCl_3$, the discoveries of half-integer quantized thermal Hall conductivity[13,14] and quantum oscillations of thermal conductivity[15] have generated great excitement for the possibility of fractional spin excitations. Critical evidence for unconventional spin excitations in $RuCl_3$ is also given by the broad spectroscopic continuum in neutron scattering[16], Raman scattering[17] and terahertz absorption[18] measurements. This continuum signifies the multiparticle excitations of fractional particles, which are distinct from the well-defined sharp peaks for magnon excitations. Although the breakdown of conventional magnons has also been proposed to describe the continuum[19], some intriguing features have been uncovered including its fermionic character[20,21] and its prominence below $T_{\mathrm{H}}$ ~ 100 K (Refs.[16–18,20,21]) where the Kitaev exchange dominates over other spin interactions.

To further identify these excitations, nonequilibrium measurements of transient dynamics have been proposed due to their superior sensitivity to excitations with distinct energy scales[22,23]. The Kramers-Kronig relations link the real part of the frequency-dependent dielectric function to an integral containing the imaginary part *over all frequencies*, and vice versa, enabling the detection of low-energy excitations in the near-infrared transient optical response. An excellent example is the measurement of quasiparticle dynamics across the superconducting gap (~ meV) in cuprate and iron-based superconductors using probe light at 1.55 eV (Refs.[24–30]). Moreover, there has also been growing interest in the nonequilibrium spin dynamics of Kitaev magnets because of the possibilities of driving the system into novel magnetic states using mechanisms such as ultrafast (de)magnetization[31,32], carrier photodoping[33,34], Floquet engineering[35,36] and quench dynamics[22,37]. Despite their profound significance, experimental studies of nonequilibrium photoexcited states in Kitaev materials are still in their early stages[31,34,38–42].

Most ultrafast work has so far focused on the low temperature regime[31,34,38,39,42], where the long-range magnetic correlation obscures the influence of the Kitaev interaction on the optical response. In $Na_2IrO_3$, a change in transient optical dynamics across the Néel temperature has been observed[38], indicating an abrupt increase in the binding energy of Hubbard excitons upon entering the magnetically orderd phase. However, direct evidence for photoexcited spin states exclusively associated with the Kitaev paramagnet has been elusive. It is therefore worthwhile to explore intriguing photoexcited spin states at intermediate temperatures, where the system is likely a proximate Kitaev paramagnet.

Here we probe the ultrafast dynamics of photoinduced excitations in single crystals of $RuCl_3$ using heterodyne transient grating spectroscopy. To determine photoexcitations of different origins, we scrutinize the complex transient reflectance change across different temperature and in-plane magnetic field regimes. Below the Néel temperature ($T_{\mathrm{N}}$ ~ 7 K), we observe photoexcitations related to the long-range antiferromagnetic order. The transient reflectance exhibits a sharp phase transition at $T_{\mathrm{N}}$ and its amplitude is suppressed under the magnetic field. Most strikingly, in the Kitaev paramagnetic temperature regime, we identify a component in the photoinduced transient reflectance that undergoes substantial amplitude shifts under the magnetic field. Additionally, the amplitude of this component fades away with increasing temperature and vanishes around $T_{\mathrm{H}}$ ~ 100 K. These observations suggest that this component is associated with unusual spin excitations in the proximate Kitaev spin liquid. This component has two lifetimes of ~ 10 ps and ~ 55 ps, which may be linked to the dynamics of exotic Kitaev spin excitations.

## Phased-resolved transient reflectance

Heterodyne transient grating spectroscopy is a powerful technique for measuring *phase-sensitive* ultrafast optical response[43] (see Supplementary Note 2 for details). It has been applied to study photoinduced nonequilibrium excitations in a variety of strongly correlated materials[38,44–46], as well as ambipolar transport and spin transport in semiconductors[47–52]. We measure the complex transient reflectance $\Delta R$, which is composed of in-phase (Re) and out-of-phase (Im) parts relative to the equilibrium reflectivity $R$ (Figure 1c inset). By inspecting both parts at relevant temperature and magnetic field regimes, we can distinguish excitations with different origins. In stark contrast, only the in-phase component is measured in conventional pump-probe reflectance experiments (see Supplementary Note 2 for details).

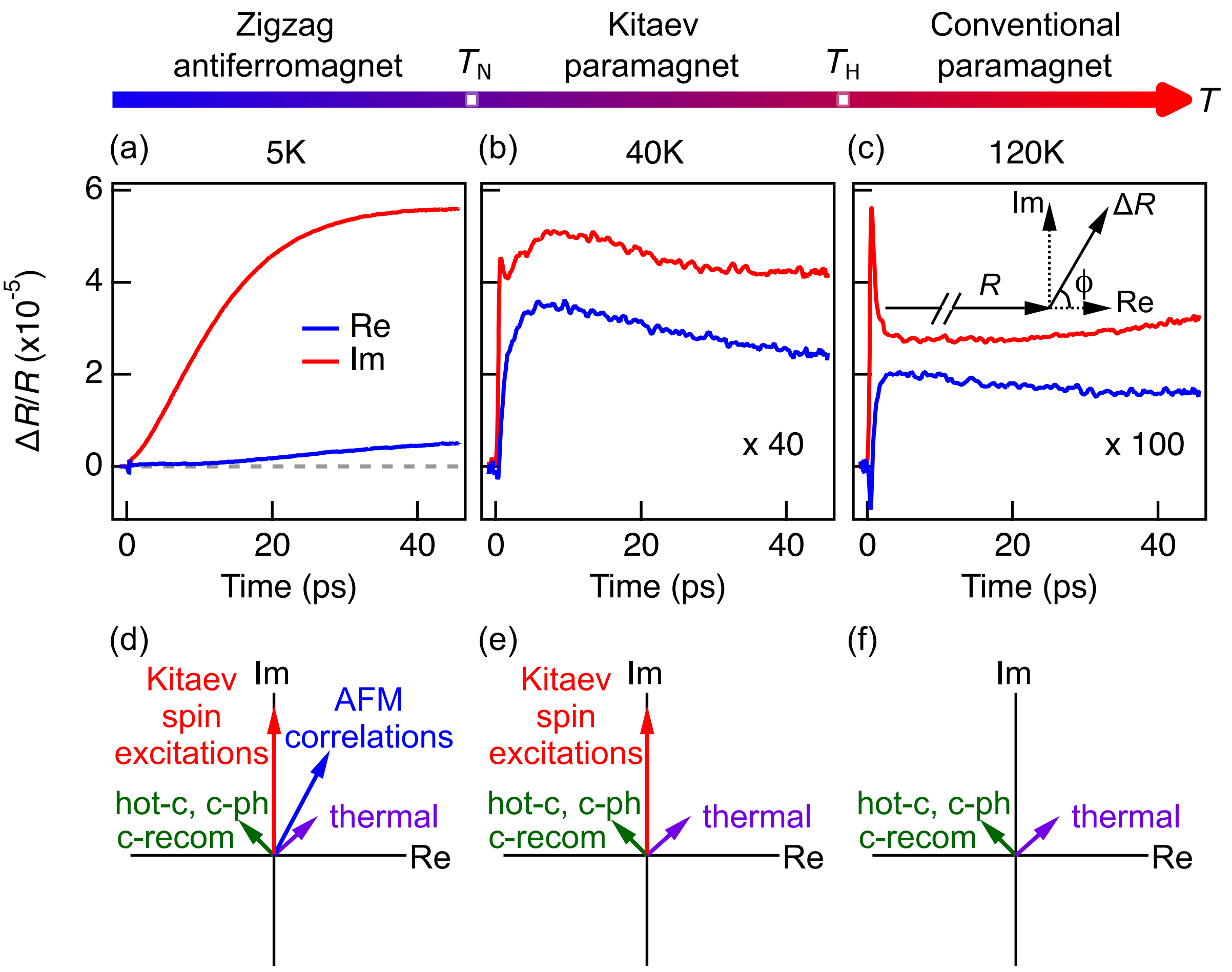

Figure 1 Transient grating signals in $RuCl_3$ at three representative temperatures under zero magnetic field with a 190 nJ cm$^{-2}$ laser fluence. (a)-(c) $\mathrm{Re}\{\Delta R/R\}$ and $\mathrm{Im}\{\Delta R/R\}$ at 5, 40 and 120 K, respectively. Note that the 40 K and 120 K data were magnified by 40 and 100 times, respectively. At 120 K (the conventional paramagnetic state), the signals reveal the cooling of photoexcited carriers, carrier recombination, carrier-phonon relaxation, and thermal processes. At 40 K (the Kitaev paramagnetic state), the in-phase part of the signal shows the same dynamics as that at 120 K, while the out-of-phase part exhibits additional contributions likely related to Kitaev spin excitations. At 5 K (the antiferromagnetic state), a new signal associated with the antiferromagnetic (AFM) correlations emerges. Inset: Diagram to illustrate the complex nature of transient reflectivity $\Delta R$. While conventional

pump-probe experiments only probe the in-phase part of $\Delta R$ (with respect to the equilibrium reflectance $R$), the transient grating spectroscopy provides information on both the in-phase and out-of-phase parts of $\Delta R$. (d)-(f) Schematics of the $\Delta R$ phase diagrams in the three different temperature regimes. hot-c: hot carriers, c-ph: carrier-phonon, c-recom: carrier recombination.

To compare the $\Delta R/R$ responses from different origins, we first present $\mathrm{Re}\{\Delta R/R\}$ and $\mathrm{Im}\{\Delta R/R\}$ as a function of time at three representative temperatures, with a laser fluence of 190 nJ cm$^{-2}$. The signals are noticeably different in the three different temperature regimes: high (Figure 1c), intermediate (Figure 1b) and low (Figure 1a).

In the high temperature regime (> 100 K), the system is in the conventional paramagnetic phase. As evident in Figure 1c, both $\mathrm{Re}\{\Delta R/R\}$ and $\mathrm{Im}\{\Delta R/R\}$ have an initial rapid decay of less than 1 ps duration and a slow decay of ~ 100 ps at 120 K. These timescales are commonly observed in semiconductors[53] and other Mott insulators[38,39]. The fast decay likely results from the cooling of photoexcited hot carriers as they reach quasi-thermal equilibrium with the lattice via carrier-carrier and carrier-phonon scattering. The slow decay reflects the carrier recombination. The ultimate return to equilibrium via the thermal diffusion takes place on a much longer timescale, shown as offsets in the figure. These processes can be attributed to two $\Delta R/R$ components with different phases, as illustrated in Figure 1f. These components are insensitive to temperature and external magnetic field, indicating that they correspond to conventional single-particle excitations.

Below $T_{\mathrm{H}}$ ~ 100 K, the system enters the Kitaev paramagnetic phase, where short-range spin correlations become strongly entangled and exotic fractionalized spin excitations may emerge. Compared with the 120 K data, $\mathrm{Re}\{\Delta R/R\}$ at 40 K shown in Figure 1b displays similar rapid and then slow decays, characteristic of the conventional pump-probe signals. $\mathrm{Re}\{\Delta R/R\}$ in this temperature regime shows weak temperature dependence (Figure S4 in the Supplementary Materials) and magnetic field dependence (Figure S5 in the Supplementary Materials). Conversely, $\mathrm{Im}\{\Delta R/R\}$ exhibits a distinct dynamic behavior compared to 120 K, characterized by a component that slowly rises and decays on ~ 10 ps timescales following the initial rapid decay. This component in $\mathrm{Im}\{\Delta R/R\}$ is attributed to the spin dynamics in the Kitaev phase, as evident by its strong temperature and magnetic field dependence (discussed later). Note that these photoinduced excitations may not be revealed in conventional pump-probe reflectance experiments where only $\mathrm{Re}\{\Delta R/R\}$ is measured.

Below $T_{\mathrm{N}}$ ~ 7 K, a zigzag antiferromagnetic order appears, affecting both the in-phase and out-of-phase parts of $\Delta R/R$. As shown in Figure 1a, the $\mathrm{Re}\{\Delta R/R\}$ signal at 5 K has a slowly rising amplitude with time, which is considerably different from the higher temperature $\mathrm{Re}\{\Delta R/R\}$ curves and is related to the long-range magnetic correlations. $\mathrm{Im}\{\Delta R/R\}$ at 5 K is also dominated by slow dynamics, demonstrating that this antiferromagnetic component has projections on both the Re and Im axes.

## Long-range magnetic order

In the following, we first examine the effects of the long-range magnetic correlation on $\Delta R/R$ by investigating the temperature and magnetic field dependence of the signal near the Néel temperature.

Figure 2a and 3a present the temperature dependence of the Re and Im parts of $\Delta R/R$ versus time from 4.5 to 15 K, respectively, at a low laser fluence of 38 nJ cm$^{-2}$ to minimize laser heating effects while maintaining sufficient signal quality. Both parts have a growing amplitude with increasing temperature below $T_{\mathrm{N}}$, but they exhibit sudden jumps from 6.5 K to 7 K, suggesting a strong impact from the long-range magnetic correlation. Moreover, although $\mathrm{Re}\{\Delta R/R\}$ and $\mathrm{Im}\{\Delta R/R\}$ have similar shapes below $T_{\mathrm{N}}$, they are not proportional to each other at a given temperature, which implies that $\Delta R/R$ contains multiple contributions. Furthermore, above $T_{\mathrm{N}}$, $\mathrm{Re}\{\Delta R/R\}$ differs markedly from $\mathrm{Im}\{\Delta R/R\}$, with a much slower rise time and a negligible amplitude above 10 K. Crucially, the fact that $\mathrm{Re}\{\Delta R/R\}$ dies away around 10 K whereas $\mathrm{Im}\{\Delta R/R\}$ persists to much higher temperatures suggests that a component in $\Delta R$ is almost aligned with the Im axis. This component reflects the Kitaev spin dynamics, which will be discussed later.

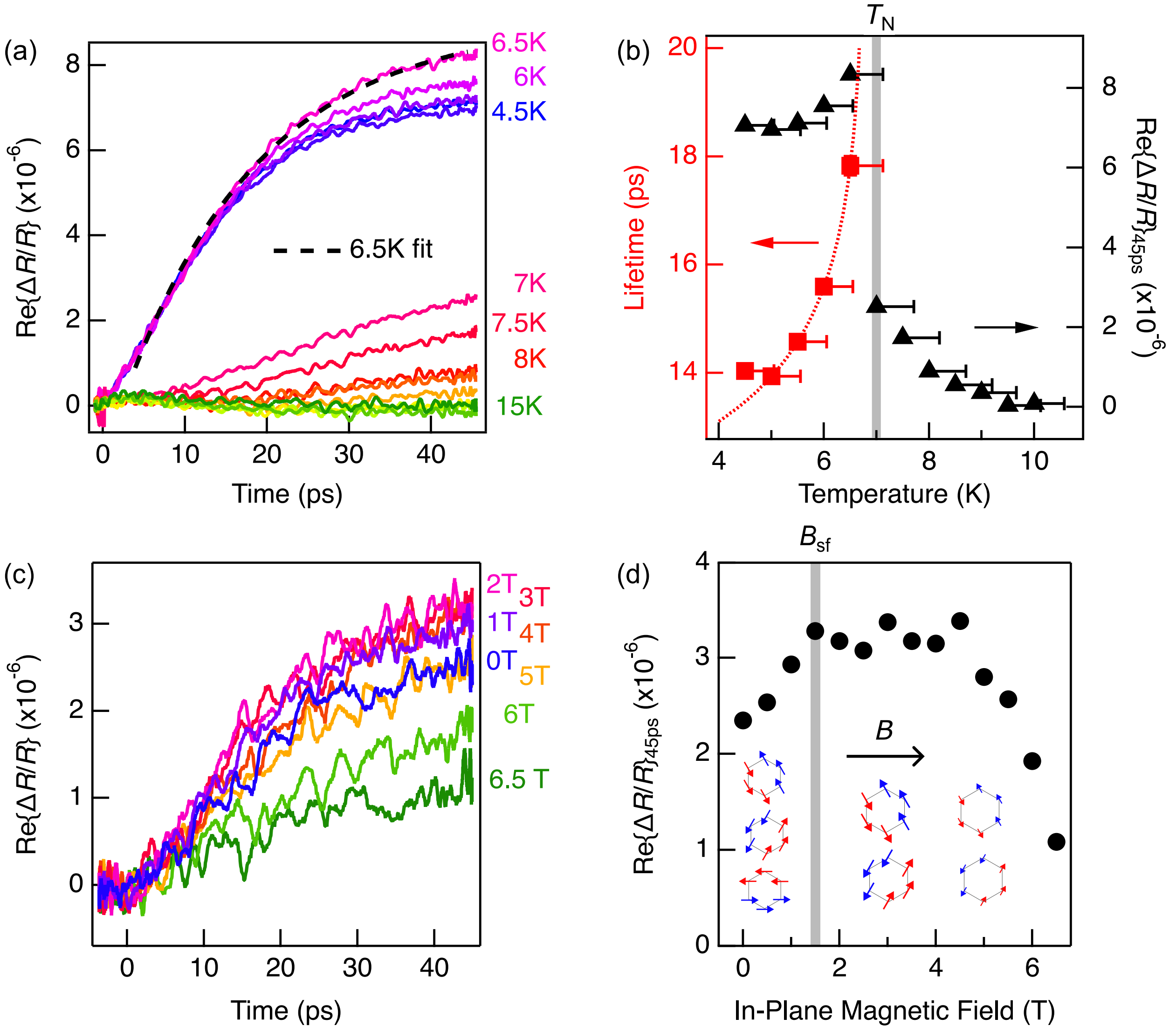

Figure 2 Effects on $\mathrm{Re}\{\Delta R/R\}$ from the antiferromagnetic order. (a) $\mathrm{Re}\{\Delta R/R\}$ from 4.5 to 15 K under zero magnetic field with a pump fluence of 38 nJ cm$^{-2}$. A gap in the signal amplitude between 6.5 K and 7 K is visible, indicating a phase transition at $T_{\mathrm{N}} \sim 7$ K. The data below $T_{\mathrm{N}}$ can be fit with an exponential function. The 6.5 K fit is shown as an example. (b) The lifetime obtained from fits to the data and the signal amplitude at 45 ps in (a) as a function of temperature. The divergence of both quantities as the temperature approaches $T_{\mathrm{N}}$ indicates a phase transition related to the long-range magnetic order. The dotted line is a power-law fit of the lifetime, showing its divergence as $T_{\mathrm{N}}$ is approached. Error bars for the lifetime are smaller than the markers when fitting the raw data in (a) to exponentials. Error bars for the temperature due to laser heating were estimated by a model in Ref.[54], using the thermal conductivity data and heat capacity data from Ref.[55] and Ref.[56], respectively. (c) $\mathrm{Re}\{\Delta R/R\}$ under an in-plane magnetic field from 0 to 6.5 T at 3.5 K with a pump fluence of 68 nJ cm$^{-2}$. The effective

temperature is lower than $T_{\mathrm{N}}$ when laser heating is considered. (d) The signal amplitude at 45 ps in (c) as a function of the magnetic field. $B_{\mathrm{sf}}$ denotes the spin-flop transition, where the three zigzag antiferromagnetic domains transition into two.

To study the long-range magnetic correlation, we focus on analyzing $\mathrm{Re}\{\Delta R/R\}$ to avoid contributions from the Im part. We fit the data below 7 K in Figure 2a with a single exponential function. We then plot the extracted lifetime and the signal amplitude at 45 ps as a function of temperature in Figure 2b. Both the lifetime and the amplitude show divergent behavior as $T_{\mathrm{N}}$ is approached, indicating that the signal is related to the critical spin fluctuations near the ordering temperature[39,57–59].

The enhancement of both the lifetime and amplitude of the photoinduced reflectivity change has been observed in various systems undergoing second-order magnetic phase transitions[39,57–60]. In particular, the zigzag antiferromagnetic order was also probed in iridates by $\mathrm{Re}\{\Delta R/R\}$ with the standard pump-probe technique[39,40]. Due to the different specific heats of electron and spin reservoirs[61], an ultrafast pump laser pulse first heats up the electrons, leaving the spin temperature unchanged. The measured lifetime reflects the timescale of the subsequent relaxation process where the spin temperature reaches quasi-equilibrium with the electron temperature[57].

Right above $T_{\mathrm{N}}$ in Figure 2a, the signal rises extremely slowly with time, implying the divergence of the spin-spin fluctuations. Based on the decreasing $\mathrm{Re}\{\Delta R/R\}$ amplitude with temperature, we determine that the critical spin fluctuations preceding the magnetic ordering disappear around 15 K.

Crucially, $\mathrm{Re}\{\Delta R/R\}$ at low temperatures is also highly sensitive to the in-plane magnetic field, providing further evidence for its magnetic origin. As shown in Figure 2d, the amplitude of $\mathrm{Re}\{\Delta R/R\}$ at 3.5 K increases up to ~ 1.5 T when the field is applied. The signal then remains on a plateau and eventually drops starting at ~ 4.5 T until the maximum field of 6.5 T. The 1.5 T transition matches well with the spin-flop transition determined by the neutron scattering[62] and magneto-optical spectroscopy[63], which corresponds to the repopulation of the three zigzag antiferromagnetic domains into two that prefer alignment with the field direction. The amplitude drop at 4.5 T agrees with previous results on the decline of the magnetic order[62,63]. Note that an in-plane field of around 6 T can induce a transition into an intermediate ordered phase due to interlayer exchange coupling (Ref.[63–65]), but this transition is unlikely to appear in our data due to the limited field strength available.

## Spin excitations in the Kitaev paramagnetic regime

As mentioned earlier, a component in $\Delta R/R$ appears only in the Im axis and below $T_{\mathrm{H}}$ ~ 100 K. We now investigate this component in detail. Figure 3a and b provides the temperature dependence of $\mathrm{Im}\{\Delta R/R\}$ from 4.5 to 15 K at a pump fluence of 38 nJ cm$^{-2}$ and from 10 to 120 K at 190 nJ cm$^{-2}$, respectively. In Figure 3b, the thermal development of this component can be clearly identified as the growing bump between 10 ps and 30 ps as the temperature decreases.

To isolate the component that is unique in the Kitaev parameter temperature regime, we subtract the contributions of the long-range magnetic correlations and the single-particle excitations from $\mathrm{Im}\{\Delta R/R\}$ (details in Supplementary Notes 3 and 4). The remaining signal in $\mathrm{Im}\{\Delta R/R\}$, denoted as the Kitaev spin excitations (KSEs) in Figure 1, can be fit with a double-exponential function: $A(\exp(-t/\tau_2) - \exp(-t/\tau_1))$, with two lifetimes $\tau_1$ and $\tau_2$ where $\tau_1 < \tau_2$.

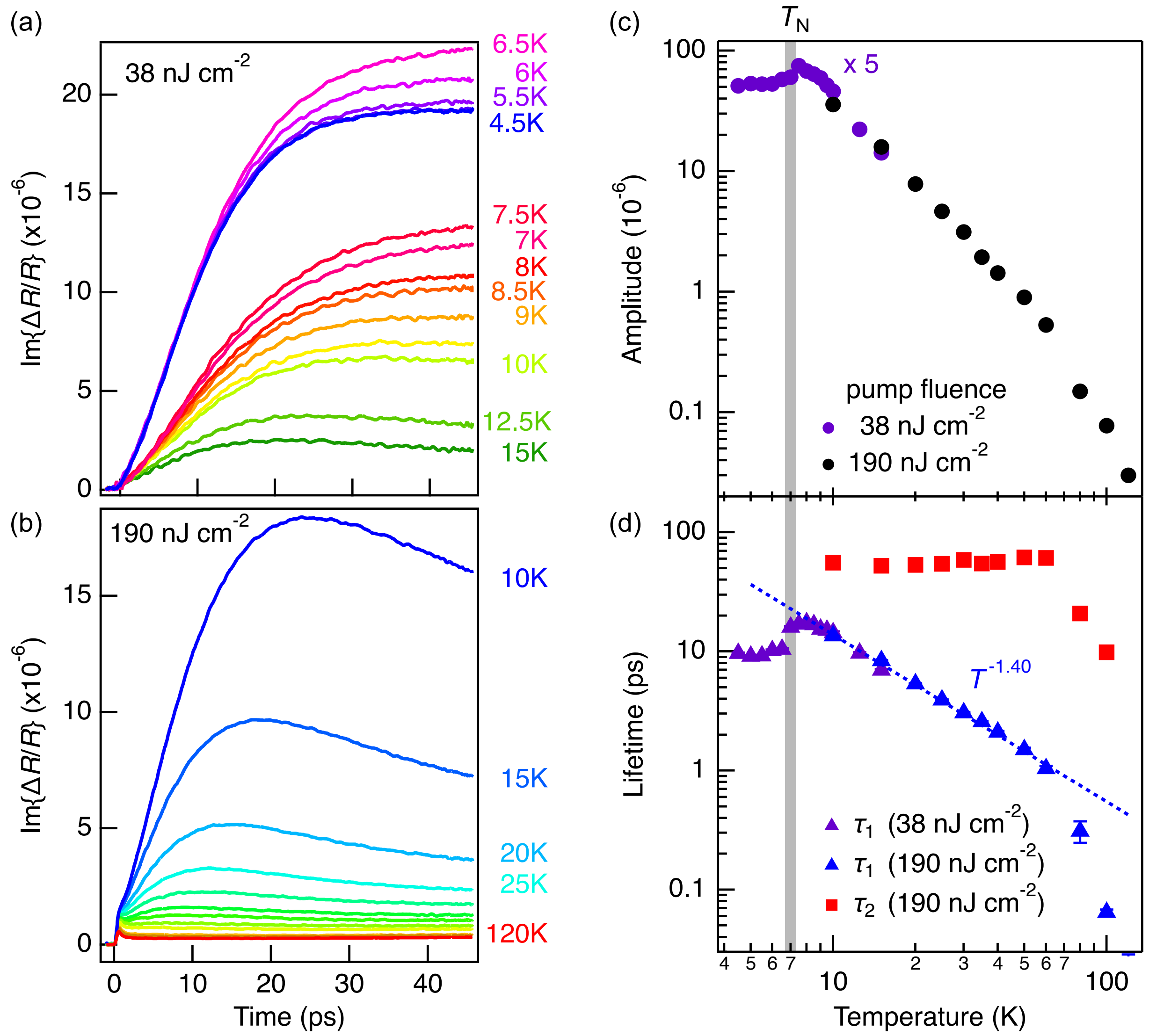

Figure 3 Temperature dependence of $\mathrm{Im}\{\Delta R/R\}$. (a)-(b) $\mathrm{Im}\{\Delta R/R\}$ from 4.5 to 15 K (38 nJ cm$^{-2}$) and from 10 to 120 K (190 nJ cm$^{-2}$). (c) Temperature dependence of the amplitude of the Kitaev spin excitations. The low fluence data is multiplied by a factor of 5 to make up for the pump power difference. (d) Temperature dependence of the lifetimes of the Kitaev spin excitations. $\tau_1$ and $\tau_2$ are the short and long lifetimes. The dotted line is the power-law fit to the data between 10 and 60 K. The temperature dependence of $\tau_1$ yields a power-law exponent of -1.40$\pm$0.05. The method to extract the amplitudes and lifetimes is described in Supplementary Notes 3 and 4. Error bars in amplitude and lifetimes represent the $\chi^2$ uncertainty when fitting the raw data in (a) and (b) to exponentials. Error bars in temperature are smaller than the markers.

Figure 3c and d shows the temperature dependence of the extracted amplitudes and lifetimes of the KSEs from the fits. The signal magnitude (and hence the extracted $A$) drops rapidly with increasing temperature, so we must increase the pump fluence at elevated temperatures to maintain acceptable signal quality. Despite the different pump fluences, the excellent overlap of the two data sets in (c) and (d) suggests that the amplitude $A$ is approximately a linear function of the pump fluence, and the lifetimes are nearly independent of the fluence in the low power regime of the experiment. Between 10 and 60 K, $\tau_1$ exhibits strong power-law temperature dependence: $\tau_1 \propto T^{-1.40}$. By contrast, $\tau_2$ (~ 55ps) is long-lived and almost temperature independent. Below 10 K, a decrease of $\tau_1$ is detected, likely owing to the development of the long-range magnetic order. Above 60 K, both the lifetimes and amplitude plummet with increasing temperature and disappear around 100 K as a result of the crossover from a Kitaev paramagnet to a conventional paramagnet, consistent with the previous terahertz, inelastic neutron and Raman scattering measurements[16–18,66,67]. Note that at 120 K this signal is negligibly small within our detection limit ($\Delta R/R \sim 3 \times 10^{-8}$), making it impossible to estimate $\tau_1$ and $\tau_2$.

In addition to the temperature dependence, the KSE component also depends strongly on the in-plane magnetic field. As shown in Figure 4a, $\mathrm{Im}\{\Delta R/R\}$ at 15 K (where there is no long-range order) shifts significantly when a magnetic field (*B*) is applied. Figure 4a also shows the contribution of the conventional single-particle excitations (SP, the green curve) to $\mathrm{Im}\{\Delta R/R\}$, which illustrates that the magnetic field modulates (rather than suppresses) the KSE's amplitude from positive to negative (compared to the SP contribution) as the field increases. Importantly, the shifts of $\mathrm{Im}\{\Delta R/R\}$ under different field values ($\Delta\mathrm{Im}\{\Delta R/R\}(B) \equiv \mathrm{Im}\{\Delta R/R\}(0\ T) - \mathrm{Im}\{\Delta R/R\}(B)$) can be scaled together and overlap well with the KSE component at zero field (15 K in Figure 4b, other temperatures in Figure S7 in the Supplementary Materials), demonstrating that the magnetic field leaves the KSE component's lifetimes unaltered. Furthermore, the magnitude of the field-induced shift, calculated by $\Delta\mathrm{Im}\{\Delta R/R\}(6.5\ \mathrm{T})$ at 45 ps time delay, decays with increasing temperature in a similar manner as the KSE's amplitude (Figure 3c) and plummets at ~ 60 K. By contrast, magnetic-field dependence of $\mathrm{Re}\{\Delta R/R\}$ is negligible in the same temperature regime (see Figure S5 in the Supplementary Materials).

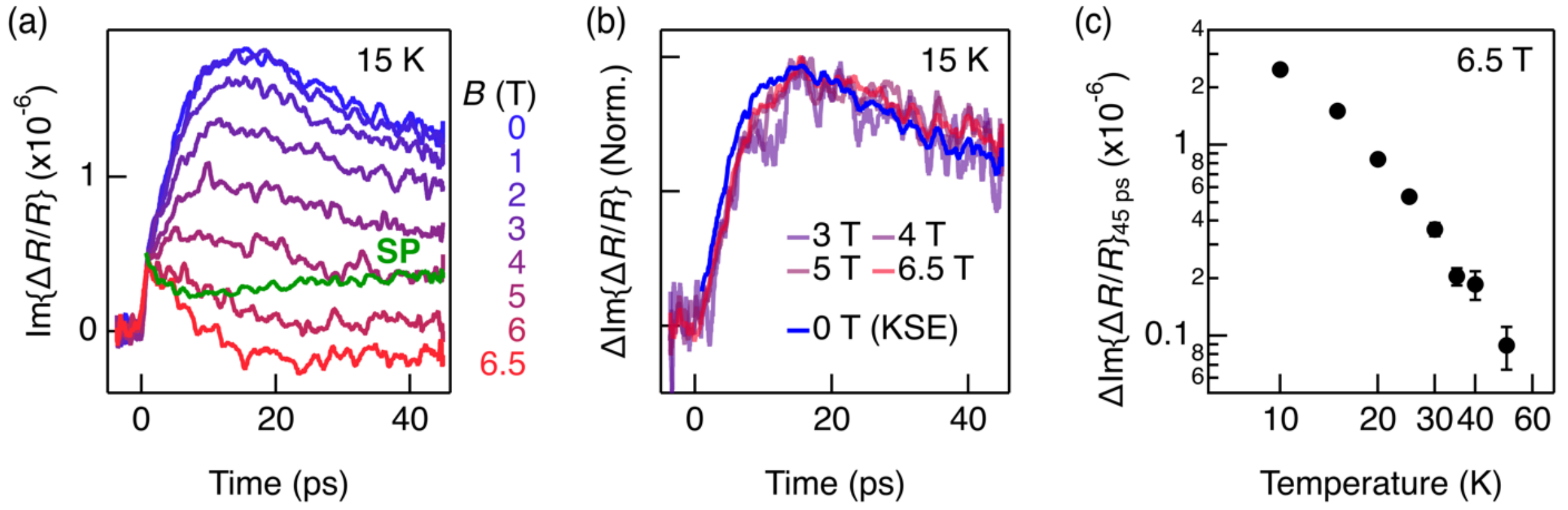

Figure 4 Magnetic field dependence of $\mathrm{Im}\{\Delta R/R\}$. (a) $\mathrm{Im}\{\Delta R/R\}$ under the magnetic field from 0 to 6.5 T at 15 K. SP (the green curve): the contribution to $\mathrm{Im}\{\Delta R/R\}$ from the conventional single-particle excitations under the zero field, inferred from the $\mathrm{Re}\{\Delta R/R\}$ signal after considering the SP component's optical phase (details in Supplementary Note 3). (b) Shift of $\mathrm{Im}\{\Delta R/R\}$, defined as $\Delta\mathrm{Im}\{\Delta R/R\}(B) \equiv \mathrm{Im}\{\Delta R/R\}(0\,T) - \mathrm{Im}\{\Delta R/R\}(B)$, from 3 to 6.5 T at 15 K. The data below 3 T are not shown due to their small signal-to-noise ratio. All data are normalized. KSE stands for the Kitaev spin excitation component in $\mathrm{Im}\{\Delta R/R\}$. (c) $\Delta\mathrm{Im}\{\Delta R/R\}(6.5\ \mathrm{T})$ amplitude at 45 ps time delay as a function of temperature. Original time traces for other temperatures are shown in Figure S6 in the Supplementary Materials. The data here is taken with a pump fluence of 517 nJ $cm^{-2}$.

The above analysis shows that the KSE component not only appears within the Kitaev paramagnetic temperature regime but also depends sensitively on the in-plane magnetic field, confirming that it originates from the magnetic photoexcitations in the Kitaev paramagnetic regime. We speculate that this component probes the spin states in the energy continuum observed in neutron scattering[16], Raman scattering[17] and terahertz absorption[18] experiments. After the pump laser pulse excites the system across the Hubbard gap, the relaxation process of photoinduced carriers would populate these low-energy spin states. It is likely that the spectral weight transfer from the optical resonance at the Hubbard bands to the low-energy spin excitations causes a significant transient optical response at the probe laser wavelength. This process is analogous to the ultrafast optical studies on cuprate and iron-based superconductors with the near infrared laser[24–30], where the photoinduced conversion from a small fraction (1%) of the superconducting condensate to the quasiparticles at the gap edges can cause a reflectance change of $10^{-4}$ at 1.55 eV (Ref.[29,68]), even if the superconducting gap is as small as 50 meV.

A recent theoretical study[22] has investigated the transient dynamics of Majorana fermions (MFs) induced by quenching an external magnetic field in the Kitaev model. It has unveiled that the itinerant MFs have a shorter lifetime than the localized MFs, both with lifetimes of ~ 10 ps. If we associated the two lifetimes in our study with the fractional particle dynamics, the itinerant MFs would have a shorter lifetime with increasing temperature due to the strong spin-phonon scattering in $RuCl_3$, while the localized MFs would have a more thermally stable scattering rate. Although this qualitatively describes our results, the calculation of the spin-phonon scattering rate from the MF self-energy shows a $T^2$ dependence[69], inconsistent with our $T^{1.40}$ result. Future theoretical work is needed to consider the relaxation and the scattering processes between the photoinduced charge, spin, and phonon excitations.

## Conclusions

In summary, we have studied the ultrafast spin dynamics in $RuCl_3$, a strong Kitaev spin liquid candidate, using optical transient grating spectroscopy. The transient complex reflectance change $\Delta R/R$ reveals different magnetic states in $RuCl_3$ in different temperature and magnetic field regimes. Near the Néel temperature, we have observed photoexcitations related to the long-range antiferromagnetic order in $\mathrm{Re}\{\Delta R/R\}$ , which exhibits a sharp phase transition at the Néel temperature and the signal amplitude is suppressed under high in-plane magnetic fields. In the Kitaev paramagnetic temperature regime, we have identified a component in $\mathrm{Im}\{\Delta R/R\}$ whose amplitude is sensitively tuned by the magnetic field and diminishes with increasing temperature. These results suggest that this component is associated with photoinduced spin excitation in the proximate Kitaev spin liquid. This component has two distinct lifetimes of ~ 10 ps and ~ 55 ps, with the former exhibiting a $T^{-1.40}$ dependence and the latter being insensitive to temperature. The measured lifetimes match the theoretical prediction of the spin decay time of the MFs in the Kitaev model under a magnetic field quench. Our results provide direct evidence for the spin dynamics uniquely connected to the Kitaev spin liquid state, and open the door for comprehensive experimental studies of ultrafast nonequilibrium phenomena in Kitaev materials.

## Acknowledgments

The work was supported by the National Key R&D Program of China (2021YFA1400100 and 2020YFA0308800), the National Natural Science Foundation of China (12074212), NSERC (RGPIN-2019-06449, RTI-2019-00809 and RGPIN-2022-04601), CIFAR, Canada Foundation for Innovation (CFI), and Ontario Research Fund. H.Z. was also supported by funds from the University of Toronto. H.-Y. K. acknowledges the support from Canada Research Chairs program.